\documentstyle[preprint,tighten,pra,aps,latexsym,amssymb]{revtex}

\newcommand{\sign}{{\mathrm{sign}}}

\newcommand{\eq}{\begin{equation}}
\newcommand{\feq}{\end{equation}}
\newcommand{\eqn}{\begin{eqnarray}}
\newcommand{\feqn}{\end{eqnarray}}
\newcommand{\arr}{\begin{eqnarray*}}
\newcommand{\farr}{\end{eqnarray*}}

\newcommand{\p}{\partial}

\newcommand{\R}{{\mathbb R}}

\newcommand{\Z}{{\mathbb Z}}

\begin{document}
\tightenlines
\draft

\def\al{\alpha}
\def\be{\beta}
\def\ga{\gamma}
\def\de{\delta}
\def\ep{\varepsilon}
\def\ze{\zeta}
\def\io{\iota}
\def\ka{\kappa}
\def\la{\lambda}
\def\roh{\varrho}
\def\si{\sigma}
\def\om{\omega}
\def\ph{\varphi}
\def\th{\theta}
\def\te{\vartheta}
\def\up{\upsilon}
\def\Ga{\Gamma}
\def\De{\Delta}
\def\La{\Lambda}
\def\Si{\Sigma}
\def\Om{\Omega}
\def\Te{\Theta}
\def\Th{\Theta}
\def\Up{\Upsilon}

\preprint{UTF 423}

\title{BPS Black Holes in Gauged N=4, D=4 Supergravity}

\author{Dietmar Klemm\footnote{email: klemm@science.unitn.it}\\ 
\vspace*{0.5cm}}

\address{Universit\`a  degli Studi di Trento,\\
Dipartimento di Fisica,\\
Via Sommarive, 14\\
38050 Povo (TN)\\
Italia\\
\vspace*{0.5cm}      
and\\ Istituto Nazionale di Fisica Nucleare,\\
Gruppo Collegato di Trento,\\ Italia}

\maketitle
\begin{abstract}
We find solutions of the bosonic sector of gauged N=4, D=4
$\mathrm{SU(2)}\times\mathrm{SU(2)}$ supergravity, which represent
dilaton black holes with toroidal or spherical event horizons.
The axion is consistently truncated, and
the gauge group is broken to $\mathrm{U(1)}\times\mathrm{U(1)}$.
The spherical black holes carry two electric and two magnetic abelian
charges, whereas the toroidal holes have vanishing magnetic charges.
The spacetime metrics are warped products, and the manifolds
turn out to be globally hyperbolic, in contrast to standard gauged
supergravity ground states. It is shown that in the toroidal case,
there are solutions preserving one quarter or one half of the supersymmetries,
while for spherical topologies all supersymmetries are broken.
In general, the toroidal BPS states represent naked singularities,
but there is also a supersymmetric black hole with vanishing
Hawking temperature. The 1/2 supersymmetric case arises for vanishing
charges and mass, and represents the known domain wall solution of the
Freedman-Schwarz model. It provides the background in which the
black holes live.
Finally, we use Chamseddine's and Volkov's
Kaluza-Klein interpretation of gauged N=4, D=4
$\mathrm{SU(2)}\times\mathrm{SU(2)}$ supergravity to lift our solutions
to ten and eleven dimensions and to consider them as solutions to
the leading order equations of motion of the string-/M-theory
effective action.
\end{abstract}


\maketitle

\section{Introduction}
Recently, gauged supergravity theories have regained considerable 
interest, which is mainly based on the AdS/CFT correspondence
\cite{malda,witten1,witten2}, providing a duality between supergravity
on AdS backgrounds and certain superconformal field theories living
on the boundary of AdS. One of these supergravity theories is the
gauged N=4, D=4 $\mathrm{SU(2)}\times\mathrm{SU(2)}$ version of
Freedman and Schwarz \cite{freedschwarz}. Lately, Chamseddine and Volkov
discovered that this model has a Kaluza-Klein interpretation
\cite{volkov,cham}, more precisely that it can be obtained from N=1
supergravity in ten dimensions by compactifying on the group
manifold $S^3 \times S^3$. This allows to lift known solutions to
ten or eleven dimensions, and thus to regard them as solutions to
the low energy effective action of string- or M-theory.
In this way, one can find new, unconventional solutions of
the higher dimensional supergravity theories.\\
Of particular interest in this context are the BPS states preserving
some amount of supersymmetry, because they allow to study
the nonperturbative sector of string theory. Furthermore, they
play an important role in giving a microscopic explanation of
the Bekenstein-Hawking entropy of some extremal black holes \cite{vafa}.\\
This motivates to look for new supersymmetric configurations in
gauged $\mathrm{SU(2)}\times\mathrm{SU(2)}$ supergravity. This model,
which we will refer to as the Freedman-Schwarz (FS) model, is characterized
by a dilaton potential which, apart from being unbounded from below,
has no critical point, and therefore the theory does not admit a
maximally symmetric AdS vacuum. Nevertheless, a stable vacuum state
has been found by Freedman and Gibbons \cite{freedgibb},
which is a product manifold
$\mathrm{AdS}_2\times \R^2$, and preserves one quarter or one half
of the supersymmetries, the latter case occuring if one of the two
gauge coupling constants vanishes.
In this "electrovac solution", the dilaton is constant;
it is stabilized by turning on an electric field, which determines the
radius of the $\mathrm{AdS}_2$ factor. (It is quite interesting to note
that later is has been shown that the Freedman-Gibbons vacuum is also
an exact solution of string theory, to all orders in $\alpha'$
\cite{antoniadis}.)
There are also other supersymmetric vacua of the Freedman-Schwarz model,
in particular the domain wall solution \cite{cowdall,singh1} preserving one
half of the supersymmetries. This solution has vanishing gauge fields,
it is purely dilatonic.
Furthermore, BPS configurations involving a non-zero axion were
found by Singh \cite{singh1,singh2}, and
non-abelian magnetic solitons were discovered by Chamseddine and Volkov
\cite{volkov,cham}. Yet, to the author's know\-ledge, until now no
BPS extremal black hole solutions have been found in the FS model, and
to remedy this will be the main purpose of the present paper
\footnote{For BPS
black holes in five-dimensional gauged supergravities coupled to
vector multiplets, cf.~\cite{behrndt1,behrndt2}}.
We will see that such BPS black holes exist only if one admits nontrivial
spacetime topology, so the interesting physics of the so-called
topological black holes \cite{amin,mann,lemos,cai,huang,vanzo}
will come into play.\\
The rest of this article is organized as follows:\\
In section \ref{sugra} the model is presented, and in \ref{bhsol}
we construct black hole solutions. Supersymmetry of these solutions
will be studied in \ref{susy}, and in \ref{lift} we shall lift
the found BPS states to ten and eleven dimensions.\\
Finally, our results are summarized and discussed in \ref{disc}.

\section{Gauged N=4, D=4 $\mathrm{SU(2)}\times\mathrm{SU(2)}$ Supergravity} 
\label{sugra}

There are two known versions of ungauged N=4, D=4 supergravity. The
first one has a global $\mathrm{SO(4)}$ symmetry \cite{das,cremmscherk}, and it
is possible to gauge the full $\mathrm{SO(4)}$
group \cite{dasfischroc}. Besides,
as was shown by Gates and Zwiebach \cite{gateszwie},
one can also gauge $\mathrm{SU(2)}\times\mathrm{SU(2)}$, leading to
another scalar potential.\\
The superalgebra of the second ungauged version \cite{cremmscherkferr}
has an $\mathrm{SU(4)}$ automorphism
group, and gauging of an $\mathrm{SU(2)}\times\mathrm{SU(2)}$
subgroup of it leads to the Freedman-Schwarz model \cite{freedschwarz}.
It is this model we are interested in, and which we shall present in the
following.\\
The theory contains one graviton $V_{\mu}^{\alpha}$\footnote{Early greek
letters refer to the tangent space, whereas late ones denote world
indices.}, four Majorana gravitinos $\psi^I_{\mu}$, six vector
fields $A^a_{\mu}$ and $B^a_{\mu}$ ($a = 1,2,3$),
four spin 1/2 fields $\chi^I$, one
dilaton $\phi$, and one axion $\eta$.
The bosonic part of the Lagrangian is \cite{freedschwarz}
\footnote{Our Lagrangian results from that in \cite{freedschwarz} by setting
$K=1$.}
\eqn
{\cal L} &=& -\frac{V}{4}R + \frac{V}{2}[\partial^{\mu}\phi\partial_{\mu}\phi
             + \exp(4\phi)\partial^{\mu}\eta\partial_{\mu}\eta]
             + \frac{\La^2}{8}V\exp(2\phi) \nonumber \\
         & & -\frac{V}{4}(A^a_{\mu\nu}A^{a\mu\nu} + B^a_{\mu\nu}B^{a\mu\nu})
             \exp(-2\phi) - \frac{V}{2}\eta(\tilde{A}^a_{\mu\nu}A^{a\mu\nu} +
             \tilde{B}^a_{\mu\nu}B^{a\mu\nu}),
\label{lagrange}
\feqn
where
\eqn
A^a_{\mu\nu} &=& \partial_{\mu}A^a_{\nu} - \partial_{\nu}A^a_{\mu}
                 + e_A \epsilon_{abc}A^b_{\mu}A^c_{\nu}, \label{Amunu} \\
B^a_{\mu\nu} &=& \partial_{\mu}B^a_{\nu} - \partial_{\nu}B^a_{\mu}
                 + e_B \epsilon_{abc}B^b_{\mu}B^c_{\nu}, \label{Bmunu} \\
\tilde{A}^a_{\mu\nu} &=& \frac{1}{2V}\epsilon_{\mu\nu}^{\;\;\;\rho\sigma}
                         A^a_{\rho\sigma}.
\feqn
$e_A$ and $e_B$ denote the gauge coupling constants, and
\eq
\La^2 = e_A^2 + e_B^2.
\feq
The supersymmetry transformations for a purely bosonic background read
\eqn
\delta\bar{\chi} &=& \frac{i}{\sqrt 2}\bar{\epsilon}(\partial_{\mu}\phi
                     +i\ga_5\exp(2\phi)\partial_{\mu}\eta)\ga^{\mu} -
                     \frac{1}{2}\exp(-\phi)\bar{\epsilon}C_{\mu\nu}
                     \sigma^{\mu\nu} + \frac{1}{4}\exp(\phi)\bar{\epsilon}
                     (e_A + i\ga_5e_B), \label{transchi} \\
\delta\bar{\psi}_{\rho} &=& \bar{\epsilon}(\overleftarrow{D}_{\rho}-\frac{i}{2}
                            \exp(2\phi)\ga_5\partial_{\rho}\eta) - \frac{i}
                            {2\sqrt 2}\exp(-\phi)\bar{\epsilon}C_{\mu\nu}
                            \ga_{\rho}\sigma^{\mu\nu} + \frac{i}{4\sqrt 2}
                            \exp(\phi)\bar{\epsilon}(e_A + i\ga_5e_B)
                            \ga_{\rho}, \label{transpsi}
\feqn
where $\epsilon = \epsilon^I$ are four Majorana spinors. The Lorentz-
and gauge covariant derivative is given by
\eq
\overleftarrow{D}_{\rho} = \overleftarrow{\partial}_{\rho} - \frac{1}{2}
                           \omega_{\rho}^{\;\al\be}\sigma_{\al\be} +
                           \frac{1}{2}e_A\al^a A_{\rho}^a +
                           \frac{1}{2}e_B\be^a B_{\rho}^a,
\feq
the $\al^a$ and $\be^a$ denoting the generators of
the (1/2,1/2) representation of
$\mathrm{SU(2)}\times\mathrm{SU(2)}$ (we shall use the matrices
given in \cite{freedschwarz}). 
Furthermore we defined
\eq
C_{\mu\nu} = \al^a A^a_{\mu\nu} + i\ga_5\be^a B^a_{\mu\nu}.
\feq
Our conventions are as follows:
The metric signature is mostly minus $(+---)$,
$\{\ga_{\mu},\ga_{\nu}\} = 2g_{\mu\nu}$,
$\sigma_{\al\be} = \frac{1}{4}[\ga_{\al},\ga_{\be}]$, $\ga_5 = i\ga^0\ga^1
\ga^2\ga^3$, so that $\ga_5^2 = 1$. We use the representation
\eqn 
\displaystyle
&\ga_0 = \left(\begin{array}{cc}
                      -1 & 0 \\
                      0 & 1
                      \end{array} \right),\qquad
&\ga_1 = \left(\begin{array}{cc}
                      0 & \sigma_1 \\
                      -\sigma_1 & 0
                      \end{array} \right), \nonumber \\
&\ga_2 = \left(\begin{array}{cc}
                      0 & \sigma_2 \\
                      -\sigma_2 & 0
                      \end{array} \right), \qquad
&\ga_3 = \left(\begin{array}{cc}
                      0 & \sigma_3 \\
                      -\sigma_3 & 0
                      \end{array} \right),
\feqn
where the two-dimensional Pauli matrices are given by
\eq
\displaystyle
\sigma_1 = \left(\begin{array}{cc}
                      0 & 1 \\
                      1 & 0
                      \end{array} \right), \qquad
\sigma_2 = \left(\begin{array}{cc}
                      0 & -i \\
                      i & 0
                      \end{array} \right), \qquad
\sigma_3 = \left(\begin{array}{cc}
                      1 & 0 \\
                      0 & -1
                      \end{array} \right).
\feq

\section{Black Hole Solutions}
\label{bhsol}

Now we are interested in purely bosonic background
solutions of the FS model,
which represent black holes. For the time being, we shall limit ourselves
to the zero axion case. To see if this is a consistent truncation, one
has to consider the axion equation of motion following from (\ref{lagrange}),
which is given by
\eq
\partial_{\mu}(V\exp(4\phi)\partial^{\mu}\eta)+\frac{V}{2}(\tilde{A}^a_{\mu\nu}
A^{a\mu\nu} + \tilde{B}^a_{\mu\nu}B^{a\mu\nu}) = 0.
\label{axequmot}
\feq
The consistency condition for zero axion therefore reads
\eq
\tilde{A}^a_{\mu\nu}A^{a\mu\nu} + \tilde{B}^a_{\mu\nu}B^{a\mu\nu} = 0.
\label{conscond1}
\feq
The remaining equations of motion for the metric, the dilaton, and the
gauge fields are, respectively
\eqn
R_{\mu\nu} - \frac{1}{2}Rg_{\mu\nu} &=& -g_{\mu\nu}
\p_{\rho}\phi\p^{\rho}\phi + 2\p_{\mu}\phi\p_{\nu}\phi
- \frac{\La^2}{4}\exp(2\phi)g_{\mu\nu} \nonumber \\
&& -2\exp(-2\phi)\left[A^a_{\mu\rho}A_{\nu}^{a\rho} +
B^a_{\mu\rho}B_{\nu}^{a\rho} - \frac{1}{4}(A_{\rho\sigma}^a
A^{a\rho\sigma} + B_{\rho\sigma}^a B^{a\rho\sigma})g_{\mu\nu}\right],
\label{eqmotmetr} \\
\frac{1}{V}\p_{\mu}(V\p^{\mu}\phi) &=& \frac{1}{2}\exp(-2\phi)
(A_{\rho\sigma}^a A^{a\rho\sigma} + B_{\rho\sigma}^a B^{a\rho\sigma})
+ \frac{\La^2}{4}\exp(2\phi), \label{eqmotdil1} \\
0 &=& \p_{\mu}(V\exp(-2\phi)A^{a\mu\nu}) = \p_{\mu}(V\exp(-2\phi)B^{a\mu\nu}).
\label{eqmotgaugef}
\feqn
Now for the metric we make the warped product ansatz 
\eq
ds^2 = f(r)dt^2 - f(r)^{-1}dr^2 - R^2(r)d\Omega^2_k, \label{ansmetric}
\feq
where $d\Omega^2_k$ denotes the metric on a two-dimensional compact
surface $\Sigma_k$,
which we assume to be of constant curvature $k$. Without loss of
generality, we can restrict ourselves to the values $k = 0, 1, -1$,
refering to a torus, a sphere, and a Riemann surface of genus $g>1$,
respectively. For $k=1$,
the ansatz (\ref{ansmetric}) has been used previously by Chan, Horne
and Mann \cite{chhoma}
to find spherical black holes in Einstein-Maxwell-dilaton
gravity, which carry either electric or magnetic charge. Later,
for purely electric charge, these
black hole solutions were generalized by Cai, Ji and Soh to the cases
$k = 0, -1$ \cite{caijisoh}, leading to the so-called topological
dilaton black holes.\\
We assume that the dilaton depends on the coordinate $r$ only,
$\phi = \phi(r)$. Furthermore we break the gauge group
$\mathrm{SU(2)}\times\mathrm{SU(2)}$ to $\mathrm{U(1)}\times\mathrm{U(1)}$,
setting
\eq
A^a_{\mu} = A_{\mu}\delta^{a3}, \qquad B^a_{\mu} = B_{\mu}\delta^{a3}.
\feq
From (\ref{Amunu}) then follows
\eq
A^a_{\mu\nu} = (\p_{\mu}A_{\nu} - \p_{\nu}A_{\mu})\delta^{a3} \equiv
               A_{\mu\nu}\delta^{a3},
\feq
with a similar equation for $B^a_{\mu\nu}$. The gauge field equations
of motion (\ref{eqmotgaugef}) are satisfied, if one sets
\eqn
A &=& \frac{Q_A}{R^2}\exp(2\phi)dt \wedge dr + H_A \epsilon, \nonumber \\
B &=& \frac{Q_B}{R^2}\exp(2\phi)dt \wedge dr + H_B \epsilon,
\feqn
where $\epsilon$ is the volume form on $\Sigma_k$, and $Q_{A,B}$,
$H_{A,B}$ are the electric and magnetic charge parameters, respectively.
The consistency condition for zero axion (\ref{conscond1}) then reads
\eq
Q_A H_A + Q_B H_B = 0. \label{conscond2}
\feq
Using the ansatz (\ref{ansmetric}) and defining
\eq
Q^2 = Q_A^2 + Q_B^2, \qquad H^2 = H_A^2 + H_B^2,
\feq
the equation of motion (\ref{eqmotdil1}) for the dilaton becomes
\eq
\frac{1}{R^2}[R^2f\phi']' + \frac{\La^2}{4}\exp(2\phi) + \frac{H^2}{R^4}
\exp(-2\phi) - \frac{Q^2}{R^4}\exp(2\phi) = 0, \label{eqmotdil2}
\feq
where the prime denotes the derivative with respect to $r$. (\ref{eqmotmetr})
yields the equations for the metric
\eqn
-\frac{f''}{2}- \frac{R'f'}{R} -\frac{2R''f}{R} &=& 2f\phi'^2 - \frac{\La^2}{4}
\exp(2\phi) - \frac{H^2}{R^4}\exp(-2\phi) - \frac{Q^2}{R^4}\exp(2\phi),
\label{rr} \\
-\frac{f''}{2} - \frac{R'f'}{R} &=& -\frac{\La^2}{4}\exp(2\phi)
- \frac{H^2}{R^4}\exp(-2\phi) - \frac{Q^2}{R^4}\exp(2\phi),
\label{tt} \\
-\frac{1}{2R^2}[(R^2)'f] + \frac{k}{R^2} &=& -\frac{\La^2}{4}\exp(2\phi)
+ \frac{H^2}{R^4}\exp(-2\phi) + \frac{Q^2}{R^4}\exp(2\phi). \label{ij}
\feqn
Subtracting (\ref{tt}) from (\ref{rr}), we get
\eq
\frac{R''}{R} = - \phi'^2. \label{Rphi}
\feq
Now we make the ansatz
\eq
R(r) = \ga r^N,
\feq
used in \cite{chhoma,caijisoh}, and obtain from (\ref{Rphi})
\eq
\phi(r) = \phi_0 \pm \sqrt{N(1-N)}\ln r. \label{dilat}
\feq
The remaining system to solve consists then of eqns.~(\ref{eqmotdil2}),
(\ref{tt}) and (\ref{ij}). It turns out that solutions exist either
for $N=0$ or for $N=1/2$. In both cases we only obtain spherical
or toroidal topologies, i.~e.~$k=1$ or $k=0$, but we do not find
Riemann surfaces with genus $g>1$. Maybe such solutions exist in
the two other gauged versions of N=4, D=4 supergravity
\cite{dasfischroc,gateszwie}, as they have other scalar potentials,
but we shall leave this issue for future investigations.\\
For $N=0$, the solutions read
\eq
H^2 = \frac{k\ga^2C}{2}, \qquad 4Q^2 = \La^2\ga^4 + \frac{2k\ga^2}{C},
\qquad f(r) = c_1 + c_2r + \frac{1}{2}(\La^2C + \frac{2k}{\ga^2})r^2,
\feq
where we defined
\eq
C \equiv \exp(2\phi_0),
\feq
and $c_{1,2}$ are integration constants. Note that the magnetic charges
are related to the curvature $k$ of $\Sigma_k$, which implies that
$H_{A,B}$ vanish in the toroidal case.
The $N=0$ solutions are product manifolds $\mathrm{AdS}_2\times S^1\times S^1$
or $\mathrm{AdS}_2\times S^2$, and hence represent the Freedman-Gibbons
solutions \cite{freedgibb}, with the $\mathrm{AdS}_2$ factor
viewed by an accelerated observer. The horizons determined by the zeros
of $f(r)$ therefore are merely acceleration horizons, and not black hole
event horizons. As these solutions have been studied extensively in
\cite{freedgibb}, we shall not consider them further, and we will instead
devote our attention to the more interesting case $N=1/2$.\\
We again get the relation
\eq
H^2 = \frac{k\ga^2C}{2}, \label{Hk}
\feq
so that $H=0$ for toroidal topology. The function $f(r)$ is given by
\eq
f(r) = \frac{2Q^2C}{\ga^4r} - m + \frac{1}{2}(\La^2C + \frac{2k}{\ga^2})r,
\label{fsol}
\feq
where $m$ is an integration constant related to the black hole's mass.
Furthermore, one has to choose the lower sign in (\ref{dilat}).\\
In general, one has an inner (Cauchy) horizon $r_-$ and an outer
(event) horizon $r_+$, with
\eq
r_{\pm} = \frac{m\ga^2 \pm \sqrt{m^2\ga^4 - 4Q^2C(\La^2C + 2k\ga^{-2})}}
          {\La^2C\ga^2 + 2k}.
\feq
For
\eq
m^2\ga^4 = 4Q^2C(\La^2C + 2k\ga^{-2}) \label{extremal}
\feq
we get an extremal black hole, whereas for
\eq
m^2\ga^4 < 4Q^2C(\La^2C + 2k\ga^{-2})
\feq
we have a naked singularity. (The scalar curvature diverges like $r^{-3}$
for $r \to 0$).
The Hawking temperature is given by
\eq
T_H = -\frac{Q^2C}{2\pi\ga^4r_+^2} + \frac{1}{8\pi}(\La^2C + 2k\ga^{-2});
\feq
it vanishes in the extremal case (\ref{extremal}). One can also compute
the Bekenstein-Hawking entropy, which reads
\eq
S_{BH} = \frac{\gamma^2 r_+}{4}, \qquad S_{BH} = \pi \gamma^2 r_+,
\feq
for $k=0$ and $k=1$ respectively. The asymptotic
behaviour of the found solutions is quite intriguing, as they are neither
asymptotically anti-de Sitter nor asymptotically flat. Let us take e.~g.~the
toroidal black hole solution
\eq
ds^2 = f(r)dt^2 - f(r)^{-1}dr^2 - \ga^2r(dx^2 + dy^2), \label{metrtorus}
\feq
where $x,y$ are coordinates on the torus (periodically identified,
$x \sim x + n$, $y \sim y + m$; $n,m \in \Z$), and
$f(r)$ is given by (\ref{fsol}) with $k=0$. For large $r$,
(\ref{metrtorus}) becomes
\eq
ds^2 = \frac{\La^2Cr}{2}dt^2 - \frac{2}{\La^2Cr}dr^2 - \ga^2r(dx^2 + dy^2),
\label{background}
\feq
which is conformal to Minkowski space with conformal factor $r$.
The manifold is globally hyperbolic (in both the toroidal and the spherical
case), in contrast to common gauged supergravity backgrounds
\footnote{Note, however, that also the non-abelian
BPS magnetic monopole solution found
in \cite{volkov} is globally hyperbolic.}. The Carter-Penrose diagrams
of our black hole spacetimes are identical to that of the
Reissner-Nordstr\"om black hole.

\section{Supersymmetric States}
\label{susy}
\subsection{The Toroidal Case}

We now want to study if among the found solutions there are BPS states,
i.~e.~states preserving some amount of supersymmetry.
To begin with, we shall deal with
the toroidal black hole (\ref{metrtorus}).
As the magnetic charges are required to vanish in this case, the
consistency condition (\ref{conscond2}) for zero axion is automatically
satisfied. Let us first consider the supertransformation (\ref{transchi})
of the spin 1/2 fields. The condition that their variation
be vanishing can be written as
\eq
\bar{\epsilon}({\cal M}_1 + \al^3{\cal M}_2 + \be^3{\cal M}_3) = 0,
\label{fermvar}
\feq
where the ${\cal M}_i$ are $4\times 4$ matrices given by
\eqn
{\cal M}_1 &=& \frac{i}{2r}\sqrt{\frac{f}{2}}\ga_1 + \frac{1}{4}
               \sqrt{\frac{C}{r}}(e_A + i\ga_5 e_B), \nonumber \\
{\cal M}_2 &=& \frac{Q_A}{\ga^2r}\sqrt{\frac{C}{r}}\sigma_{01}, \\
{\cal M}_3 &=& \frac{Q_B}{\ga^2r}\sqrt{\frac{C}{r}}i\ga_5\sigma_{01}. \nonumber
\feqn
Using the representation of $\al^3$, $\be^3$ given in \cite{freedschwarz},
one can write (\ref{fermvar}) in the form
\eq
\bar{\epsilon}\Theta = 0,
\feq
the $16\times 16$ matrix $\Theta$ being defined as
\eq
\Theta = \left(\begin{array}{cc}
                      \Theta_+ & 0 \\
                      0 & \Theta_-
                      \end{array} \right), \qquad
\Theta_{\pm} = \left(\begin{array}{cc}
                      {\cal M}_1 & {\cal M}_2 \pm {\cal M}_3 \\
                      -({\cal M}_2 \pm {\cal M}_3) & {\cal M}_1
                      \end{array} \right).
\feq
The necessary condition for the existence of Killing spinors $\bar{\epsilon}$
is thus $\det\Theta = 0$, or, equivalently
\eq
\det\Theta_+ = 0 \qquad \vee \qquad \det\Theta_- = 0.
\feq
The determinants read
\eq
\det\Theta_{\pm} = \frac{1}{(8\ga^2r^2)^4}\left(m^2\ga^4 - 4C^2(e_B Q_A \mp
                   e_A Q_B)^2\right)^2,
\feq
which yields
\eq
m^2\ga^4 = 4C^2(e_B Q_A \mp e_A Q_B)^2. \label{bog}
\feq
As we have the inequality
\eq
(e_B Q_A \mp e_A Q_B)^2 \le \La^2 Q^2, \label{inequ}
\feq
the configurations satisfying (\ref{bog}) are in general naked
singularities, except for the one with
\eq
e_A Q_A \pm e_B Q_B = 0, \label{satur}
\feq
which saturates (\ref{inequ}) and therefore represents an extremal
black hole.\\
Without loss of generality we shall restrict ourselves to the upper
sign in the foregoing equations, so that
\eq
\det\Theta_+ = 0, \qquad \det\Theta_- \neq 0.
\feq
This then implies the constraint
\eq
\bar{\epsilon}(\al^3-\be^3) = 0 \label{constreps}
\feq
on the Killing spinors.\\
Next we also have to impose that the gravitino variation (\ref{transpsi})
be vanishing. Using the fact that $\delta\bar{\chi} = 0$, this condition
can be simplified to
\eq
0 = \bar{\epsilon}\overleftarrow{\nabla}_{\mu}, \label{deltapsivan}
\feq
where the supercovariant derivative $\overleftarrow{\nabla}_{\mu}$
is defined by
\eq
\overleftarrow{\nabla}_{\mu} =
\overleftarrow{D}_{\mu} - \frac{i}{\sqrt{2}}\exp(-\phi)
C_{\mu\rho}\ga^{\rho} + \frac{1}{2}(\p_{\rho}\phi)\ga^{\rho}\ga_{\mu},
\label{supercovder}
\feq
whose components read
\eqn
\overleftarrow{\nabla}_t &=& \overleftarrow{\p}_t + \frac{1}{4}\left(f' -
                             \frac{f}{r}\right)\ga_0\ga_1 +
                             \al^3\left(\frac{e_A Q_A C}
                             {2\ga^2r} + \frac{iQ_A}{\ga^2}\sqrt{\frac{fC}
                             {2r^3}}\ga_1\right) \nonumber \\
                         & & \be^3\left(\frac{e_B Q_B C}{2\ga^2r} + \frac{iQ_B}
                             {\ga^2}\sqrt{\frac{fC}
                             {2r^3}}i\ga_5\ga_1\right), \\
\overleftarrow{\nabla}_r &=& \overleftarrow{\p}_r - \frac{1}{4r} +
                             \al^3\frac{iQ_A}{\ga^2}\sqrt{\frac{C}
                             {2fr^3}}\ga_0 + \be^3\frac{iQ_B}{\ga^2}
                             \sqrt{\frac{C}{2fr^3}}i\ga_5\ga_0, 
                             \label{nablar} \\
\overleftarrow{\nabla}_x &=& \overleftarrow{\p}_x, \\
\overleftarrow{\nabla}_y &=& \overleftarrow{\p}_y.
\feqn
(\ref{deltapsivan}) thus implies that Killing spinors, if they exist,
are independent of the coordinates $x,y$, and therefore they automatically
respect the identifications on the torus.\\
As no $x,y$-dependence occurs in $\overleftarrow{\nabla}_t$
and $\overleftarrow{\nabla}_r$,
the only integrability condition following from (\ref{deltapsivan})
reads
\eq
\bar{\epsilon}[\overleftarrow{\nabla}_t,\overleftarrow{\nabla}_r] = 0.
\label{integrab}
\feq
One finds
\eq
[\overleftarrow{\nabla}_t,\overleftarrow{\nabla}_r] = 
-(\overleftarrow{\nabla}_t - \overleftarrow{\p}_t)\frac{1}{r},
\feq
and hence the integrability condition (\ref{integrab}) together with
\eq
\bar{\epsilon}\overleftarrow{\nabla}_t = 0
\feq
implies that the Killing spinors are also independent of $t$,
\eq
\bar{\epsilon}\overleftarrow{\p}_t = 0.
\feq
Next we consider condition (\ref{fermvar}) in more detail, in particular
we will show that (\ref{integrab}) follows completely from $\delta\bar{\chi}
= 0$. To this aim, we define the idempotents
\eqn
\Ga_1 &=& \al^3\frac{Q_A + i\ga_5 Q_B}{iQ}\ga_0, \nonumber \\
\Ga_2 &=& i\al^3\ga_0\ga_1, \\
\Ga_3 &=& \frac{e_A + i\ga_5 e_B}{i\La}\ga_1. \nonumber \\
\feqn
Note that one has
\eqn
[\Ga_1,\Ga_3] &=& \frac{2i(e_A Q_A + e_B Q_B)}{\La Q}\Ga_2, \nonumber \\
\{\Ga_1,\Ga_2\} &=& 0.
\feqn
Using these definitions and the constraint (\ref{constreps}), the equation
$\delta\bar{\chi} = 0$ reads
\eq
\bar{\epsilon}\left[-\frac{i}{2r}\sqrt{\frac{f}{2}} - \frac{iQ}{2\ga^2r}
\sqrt{\frac{C}{r}}\Ga_1 + \frac{i\La}{4}\sqrt{\frac{C}{r}}\Ga_3\right] = 0.
\label{fermvarman}
\feq
Consider now the identities
\eqn
\left[-\frac{i}{2r}\sqrt{\frac{f}{2}}\right. &-& \left.\frac{iQ}{2\ga^2r}
\sqrt{\frac{C}{r}}\Ga_1 + \frac{i\La}{4}\sqrt{\frac{C}{r}}\Ga_3\right]
\left[-\frac{i}{2r}\sqrt{\frac{f}{2}} - \frac{iQ}{2\ga^2r}
\sqrt{\frac{C}{r}}\Ga_1 - \frac{i\La}{4}\sqrt{\frac{C}{r}}\Ga_3\right] =
\nonumber \\
& & \left(-\frac{CQ^2}{2\ga^4r^3} + \frac{m}{8r^2}\right)[1 + x(r)\Ga_1
    + y(r)\Ga_2], \label{firstid}
\feqn
and
\eqn
\left[-\frac{i}{2r}\sqrt{\frac{f}{2}}\right. &-& \left.\frac{iQ}{2\ga^2r}
\sqrt{\frac{C}{r}}\Ga_1 + \frac{i\La}{4}\sqrt{\frac{C}{r}}\Ga_3\right]
\left[-\frac{i}{2r}\sqrt{\frac{f}{2}} + \frac{iQ}{2\ga^2r}
\sqrt{\frac{C}{r}}\Ga_1 - \frac{i\La}{4}\sqrt{\frac{C}{r}}\Ga_3\right] =
\nonumber \\
& & \frac{m}{4r^2}P(\ga_5 \Ga_2\sign(e_A Q_B - e_B Q_A)), \label{secondid}
\feqn
where
\eqn
x(r) &=& \frac{2\sqrt{2frC}Q\ga^2}{4CQ^2 - \ga^4mr}, \nonumber \\
y(r) &=& \frac{2iC(e_A Q_A + e_B Q_B)\ga^2r}{4CQ^2 - \ga^4mr},
\feqn
obeying
\eq
x^2 + y^2 = 1. \label{xy1}
\feq
Furthermore we defined
\eq
P({\cal L}) = \frac{1}{2}(1 + {\cal L}),
\feq
for an idempotent operator ${\cal L}$ ($P({\cal L})$ is then a projector).
(\ref{xy1}) implies that also $(1+x\Ga_1+y\Ga_2)/2$ is an ($r$-dependent)
projection operator. One easily shows that
\eq
[(1+x\Ga_1+y\Ga_2)/2,P(\ga_5 \Ga_2\sign(e_A Q_B - e_B Q_A))] = 0.
\feq
Now, using the identities (\ref{firstid}) and (\ref{secondid}), one finds that
(\ref{fermvarman}) implies the two equations
\eqn
\bar{\epsilon}[1 + x(r)\Ga_1 + y(r)\Ga_2] &=& 0, \label{integrabmod} \\
\bar{\epsilon}P(\ga_5 \Ga_2\sign(e_A Q_B - e_B Q_A)) &=& 0. \label{addcond}
\feqn
It is easy to show that eq.~(\ref{integrabmod}) is equivalent to the
(\ref{integrab}), so the integrability condition follows entirely
from $\delta\bar{\chi} = 0$.
Since the operator acting on $\bar{\epsilon}$ in (\ref{fermvarman}) is
itself a linear combination of the two operators occuring in
(\ref{integrabmod}) respectively (\ref{addcond}),
\eqn
&&\left[-\frac{i}{2r}\sqrt{\frac{f}{2}} - \frac{iQ}{2\ga^2r}
\sqrt{\frac{C}{r}}\Ga_1 + \frac{i\La}{4}\sqrt{\frac{C}{r}}\Ga_3\right] =
\nonumber \\
&&\frac{1}{2i}\left(\frac{Q}{\ga^2r}\sqrt{\frac{C}{r}} -
\frac{m\ga^2}{4Q\sqrt{Cr}}\right)[1 + x(r)\Ga_1 + y(r)\Ga_2]\Ga_1 \nonumber \\
&& + \frac{m\ga^2}{4iQ\sqrt{Cr}}P(\ga_5 \Ga_2\sign(e_A Q_B - e_B Q_A))\Ga_1,
\feqn
the equation $\delta\bar{\chi} = 0$ is equivalent to the two conditions
(\ref{integrabmod}) and (\ref{addcond}). Thus the only remaining equations
to solve are these constraints, and the radial equation
\eq
\bar{\epsilon}\overleftarrow{\nabla}_r = 0, \label{radequ}
\feq
with $\overleftarrow{\nabla}_r$ given by (\ref{nablar}).
Eq.~(\ref{radequ}) is of the form
\eq
\bar{\epsilon}\left[\overleftarrow{\p}_r - a(r) - b(r)\Ga_1\right] = 0,
\feq
where we did not write down the functions $a(r)$ and $b(r)$ explicitely.
The solution of a spinorial equation of this type, subject to constraints
like (\ref{integrabmod}) and (\ref{addcond}),
is given in the appendix of \cite{romans},
and reads
\eq
\bar{\epsilon} = \bar{\epsilon}_0 P(-\ga_5 \Ga_2\sign(e_A Q_B - e_B Q_A))
                 P(-\Ga_1)(u(r) + v(r)\Ga_2),
\feq
with
\eqn
u(r) &=& \sqrt{4CQ^2 - \ga^4mr + 2\sqrt{2frC}Q\ga^2}r^{-1/4}, \nonumber \\
v(r) &=& -\sqrt{4CQ^2 - \ga^4mr - 2\sqrt{2frC}Q\ga^2}r^{-1/4},
\feqn
and $\bar{\epsilon}_0$ denoting a constant spinor.
Finally, we have to implement also the condition (\ref{constreps}),
which is equivalent to
\eq
\bar{\epsilon}P(\be^3\al^3) = 0.
\feq
This yields for the final version of the Killing spinors
\eq
\bar{\epsilon} = \bar{\epsilon}_0 P(-\be^3\al^3)
                 P(-\ga_5 \Ga_2\sign(e_A Q_B - e_B Q_A))
                 P(-\Ga_1)(u(r) + v(r)\Ga_2). \label{killing}
\feq
In the case of an extremal black hole, we have
\eq
e_A Q_A + e_B Q_B = 0,
\feq
(cf.~(\ref{satur})), and the Bogomol'nyi bound (\ref{bog}) reads
\eq
m^2\ga^4 = 4Q^2C^2\La^2.
\feq
The Killing spinors (\ref{killing}) then simplify to
\eq
\bar{\epsilon} = \bar{\epsilon}_0 P(-\be^3\al^3)
                 P(-\ga_5 \Ga_2\sign(e_A Q_B - e_B Q_A))
                 P(\Ga_1)f(r)^{1/4}. \label{killingextr}
\feq
The three projectors appearing in (\ref{killing}) and (\ref{killingextr})
reduce the dimension of the solution space from sixteen to two, enough for
a residual N=1 supersymmetry. For vanishing electric charges, $Q=0$,
(\ref{bog}) implies $m=0$, in which case the metric reduces to
(\ref{background}). This space represents the background of our black
hole solutions.
Then $\delta\bar{\chi} = 0$ takes the simple
form
\eq
\bar{\epsilon}P(\Ga_3) = 0,
\feq
i.~e.~in the uncharged case it is not necessary
to decompose $\delta\bar{\chi} = 0$ into
two independent projection conditions, as we were forced to do
for the charged configurations.
The radial equation (\ref{radequ}) reads
\eq
\bar{\epsilon}\left[\overleftarrow{\p}_r - \frac{1}{4r}\right] = 0.
\feq
This yields for the Killing spinors
\eq
\bar{\epsilon} = \bar{\epsilon}_0 P(-\Ga_3)r^{1/4}.
\feq
The number of Killing spinors is thus enhanced from two to eight,
corresponding to an N=2 supersymmetry algebra, which means
that the background preserves one half of the supersymmetries.
This background is identical to the domain wall solution found in
\cite{cowdall,singh1}.\\
Another special case occurs if
\eq
e_A Q_B \mp e_B Q_A = 0,
\feq
for which we also have $m=0$ from (\ref{bog}). Similarly to the situation
for $Q=0$ desribed above, $\delta\bar{\chi} = 0$ is itself a projection
condition, doubling this time the number of linearly independent Killing
spinors from two to four\footnote{Note that e.~g.~for $e_A Q_B - e_B Q_A =
m = 0$, one has $\det\Theta_+ = 0$, and $\det\Theta_- \neq 0$, so one still
needs the constraint (\ref{constreps}), whereas for $Q = m = 0$ we have
$\det\Theta_+ = \det\Theta_- = 0$, so (\ref{constreps}) can be dropped.},
which corresponds to a residual N=1 supersymmetry.

\subsection{The Spherical Case}

Let us now consider the configurations with topology $\R^2 \times S^2$.
The spacetime metric is then
\eq
ds^2 = f(r)dt^2 - f(r)^{-1}dr^2 - \ga^2r(d\theta^2 + \sin^2\theta d\varphi^2),
\feq
where $f(r)$ is given by (\ref{fsol}) with $k=1$.\\
Again we first require that the
variation of the spin $1/2$ fields be vanishing. One can then
proceed like in the toroidal case, and gets for the determinants
$\det\Theta_{\pm}$
\eqn
\det\Theta_{\pm} &=& \frac{1}{(8\ga^4r^2)^4}\left[-(m\ga^4 \pm 
                     4(H_B Q_A - H_A Q_B))^2 + 16H^4(Q_A e_B \mp Q_B e_A)^2
                     \right. \nonumber \\
                 & & \left.
                     + 16\ga^2r(Q_A e_B \mp Q_B e_A)(H_A e_A \pm H_B e_B)H^2
                     + 4\ga^4r^2(H_A e_A \pm H_B e_B)^2\right]^2.
\feqn
These determinants must vanish identically as functions of $r$, which yields
\eqn
(m\ga^4 \pm 4(H_B Q_A - H_A Q_B))^2 &=& 16H^4(Q_A e_B \mp Q_B e_A)^2, \\
H_A e_A \pm H_B e_B &=& 0. \label{cond2}
\feqn
Again, without loss of generality we may restrict ourselves to the upper
sign in the previous equations. Then we have
\eq
\det\Theta_+ = 0, \qquad \det\Theta_- \neq 0,
\feq
yielding the constraint (\ref{constreps}).
Now, one of the various integrability conditions for Killing spinors,
following from (\ref{deltapsivan}), reads
\eq
\bar{\epsilon}
[\overleftarrow{\nabla}_{\theta},\overleftarrow{\nabla}_{\varphi}] = 0,
\feq
where the supercovariant derivatives are determined by
(\ref{supercovder}). Using (\ref{constreps}), (\ref{Hk})
and (\ref{cond2}), one gets
\eq
\bar{\epsilon}
[\overleftarrow{\nabla}_{\theta},\overleftarrow{\nabla}_{\varphi}] =
\bar{\epsilon}\al^3\frac{i\cos\theta}{\sqrt{2C}\ga}(H_A + i\ga_5 H_B)\ga_2.
\label{integrthetaphi}
\feq
Now the determinant of the matrix acting on $\bar{\epsilon}$ in
(\ref{integrthetaphi}) is zero iff the magnetic charges $H_A$ and $H_B$
vanish, i.~e.~$H=0$. This is, however, incompatible with (\ref{Hk}).
This means that among the spherical solutions we have found, there is
no configuration preserving some amount of supersymmetry. Note that this
behaviour is similar to that of the Freedman-Gibbons solutions
\cite{freedgibb}, where only the $\mathrm{AdS}_2\times \R^2$
configurations (or $\mathrm{AdS}_2\times S^1 \times S^1$, if one compactifies
the real plane to a torus) can yield BPS states,
whereas the $\mathrm{AdS}_2\times S^2$ solutions break all supersymmetries.

\subsection{Nonvanishing Axion}

We now want to give a short comment on the issue
in which way our considerations can be generalized to
the case of nonzero axion. It is well-known that the ungauged version
of N=4, D=4 supergravity considered in \cite{cremmscherkferr} has,
apart from the $\mathrm{SU(4)}$ invariance group, an additional
$\mathrm{SU(1,1)}$ symmetry of the equations of motion. Of this symmetry,
however, only an $\mathrm{U(1)}$ subgroup survives in the gauged version,
namely the group of transformations \cite{freedschwarz}
\eq
\eta(x) \to \eta(x) + c, \label{etatransf}
\feq
$c$ being a real constant\footnote{Actually, (\ref{etatransf}) is a
global invariance of the action, the additional term appearing
due to $c$ being a total derivative because of the Bianchi identities
$\p_{\mu}(V\tilde{A}^{\mu\nu}) = \p_{\mu}(V\tilde{B}^{\mu\nu}) = 0$.}.
As only the derivatives of $\eta(x)$ enter the
supersymmetry transformation laws $(\ref{transchi})$ and $(\ref{transpsi})$,
this represents a rather trivial generalization; it means that the
BPS states found above remain supersymmetric if the axion, instead of being
zero, assumes a constant value. (Note that also the axionic equation of
motion (\ref{axequmot}) remains satisfied, provided the consistency condition
(\ref{conscond1}) is fulfilled). Hence there seems to be no way of using
duality symmetries to create nontrivial axionic solutions in an efficient
manner, so we shall leave the search for the latter for future
investigations.

\section{Interpretation as BPS String- and M-Theory Solutions}
\label{lift}

Finally, we would like to use the Kaluza-Klein interpretation of gauged
N=4, D=4 $\mathrm{SU(2)}\times\mathrm{SU(2)}$ supergravity given by
Chamseddine and Volkov \cite{volkov,cham}, to lift our BPS configurations
to ten and eleven dimensions. In doing so, we could
regard them as supersymmetric
solutions of the low energy effective action of string- or M-theory.
The details of the compactification of N=1 supergravity in ten dimensions
on the group manifold $S^3 \times S^3$ are given in \cite{volkov},
so we shall not repeat them here.
We adopt the index convention used in \cite{volkov}, i.~e.~greek, latin,
and capital latin indices refer to the four-dimensional, internal
six-dimensional ($S^3 \times S^3$), and ten-dimensional spaces,
respectively. Early letters relate to the tangent space, whereas 
the late ones denote world indices. Furthermore we parametrize the
six internal coordinates by a pair of indices, that is $\{m\} = \{(s),i\}$,
with $s = 1,2$ (specifying the three-sphere)
and $i = 1,2,3$, and similarly for the tangent space,
$\{{\mathrm{a}}\} = \{(s), a\}$, where $a = 1,2,3$.
Performing the inverse of the dimensional reduction desribed in
\cite{volkov}, we arrive at the ten-dimensional metric
(in Einstein-frame)\footnote{Our
conventions differ from those used in \cite{volkov} by $\phi \to -\phi$.}
\eq
d\hat{s}^2 = \exp(3\phi/2)ds^2 - 2\exp(-\phi/2)(\Theta^{(1)a} \Theta^{(1)a}
             + \Theta^{(2)a} \Theta^{(2)a}),
\label{10dmetrEframe}
\feq
where $ds^2$ denotes the four-dimensional line element, and the forms
$\Theta^{(s)a}$ are defined by
\eqn
\Theta^{(1)a} &=& A^a + \frac{\theta^{(1)a}}{e_A}, \nonumber \\
\Theta^{(2)a} &=& B^a + \frac{\theta^{(2)a}}{e_B}.
\feqn
Here $A^a$ and $B^a$ are the gauge potentials occuring in (\ref{Amunu}) and
(\ref{Bmunu}), and the $\theta^{(s)a}$ denote the Maurer-Cartan forms on
$S^3$ satisfying
\eq
d\theta^{(s)a} + \frac{1}{2}\epsilon_{abc}\theta^{(s)b}\wedge\theta^{(s)c} = 0.
\feq
Explicitely, they read
\eqn
\theta^1 &=& \cos\psi d\vartheta + \sin\psi\sin\vartheta d\varphi, \nonumber \\
\theta^2 &=& -\sin\psi d\vartheta + \cos\psi\sin\vartheta d\varphi, \\
\theta^3 &=& d\psi + \cos\vartheta d\varphi, \nonumber
\feqn
$\psi,\vartheta,\varphi$ being the Euler angles on the three-sphere.\\
Finally, we also need to give the ten-dimensional dilaton $\hat{\phi}$, and
the antisymmetric tensor field $\hat{H}_{MNP}$ appearing in N=1, D=10
supergravity. For the former, we obtain
\eq
\hat{\phi} = \frac{\phi}{2} = \frac{\phi_0}{2} - \frac{1}{4}\ln r,
\feq
while the non-vanishing components of the latter read
\eqn
\hat{H}_{\al\be\mathrm{a}} &=& -\frac{1}{\sqrt{2}}\exp(-5\phi/4)
                               F^{\mathrm{a}}_{\al\be}, \nonumber \\
\hat{H}_{\mathrm{abc}} &=& \frac{1}{2\sqrt{2}}\exp(3\phi/4)f_{\mathrm{abc}},
\feqn
where we defined
\eq
F^{(1)a}_{\al\be} = A^a_{\al\be}, \qquad F^{(2)a}_{\al\be} = B^a_{\al\be},
\feq
and the $f_{\mathrm{abc}}$ are the $\mathrm{SU(2)}$ gauge group structure
constants,
\eq
f_{\mathrm{abc}} \equiv f^{(s)}_{abc} = e_s\epsilon_{abc}, \qquad s=A,B.
\feq
The ten-dimensional metric (\ref{10dmetrEframe}) has a rather complicated
form, in particular we note that the four-dimensional gauge fields now
give rise to off-diagonal components of the metric. (\ref{10dmetrEframe})
is different from known brane-like solutions of ten-dimensional
supergravity, and it also does not seem to be an intersection of branes.
However, asymptotically, i.~e.~for $r \to \infty$, the metric approaches
a simple form. To see this, we write the line element (\ref{10dmetrEframe})
in string frame,
\eq
d\tilde{s}^2 = \frac{1}{2}\exp(\hat{\phi})d\hat{s}^2 =
               \frac{1}{2}\exp(2\phi)ds^2 - \Theta^{(1)a} \Theta^{(1)a} -
               \Theta^{(2)a} \Theta^{(2)a}.
\label{10dmetrSframe}
\feq
Now introduce the new coordinates
\eq
\tau = \frac{\La C}{2}t, \qquad \rho = \frac{\ln r}{\La}, \qquad
X = \sqrt{\frac{C}{2}}\ga x, \qquad Y = \sqrt{\frac{C}{2}}\ga y.
\feq
This yields for the metric (\ref{10dmetrSframe}) in the limit
$r \to \infty$
\eq
d\tilde{s}^2 = d\tau^2 - d\rho^2 - dX^2 - dY^2 - \frac{1}{e_A^2}
               \theta^{(1)a}\theta^{(1)a} - \frac{1}{e_B^2}
               \theta^{(2)a}\theta^{(2)a}. \label{intersNS5}
\feq
Asymptotically, our BPS configurations thus approach (in string frame)
$M^2 \times S^1 \times S^1 \times S^3 \times S^3$,
where $M^2$ denotes
two-dimensional Minkowski space, and the radii of the three-spheres are given
by $1/e_A$ and $1/e_B$. (\ref{intersNS5}) arises also for vanishing gauge
fields and mass parameter. If we do not consider the $X,Y$-directions
as compactified, we recognize (\ref{intersNS5}) to be the
near-horizon limit of the intersection
of two NS five-branes on a line. This supersymmetric solution
of N=1, D=10 supergravity has previously been considered in \cite{cowtown}.
Upon compactification on $S^3 \times S^3$, it gives rise to the
domain wall solution of the FS model.\\
One can further lift the solutions to eleven dimensions, to regard them in
the context of M-theory. Using the rules of \cite{volkov}, one obtains
for the metric
\eq
ds^2_{(11)} = \exp(4\phi/3)ds^2 - 2\exp(-2\phi/3)(\Theta^{(1)a} \Theta^{(1)a}
              + \Theta^{(2)a} \Theta^{(2)a}) - \exp(4\phi/3)dx^{(10)}dx^{(10)},
\feq
and for the antisymmetric tensor field
\eq
A_{MN10} = C_{MN}, \qquad A_{MNP} = 0,
\feq
where $C_{MN}$ is the two-form potential appearing in ten dimensions,
which gives rise to the field strength $H_{MNP}$ via $H = dC$.

\section{Conclusions}
\label{disc}

To sum up, we have obtained several new multi-parameter solutions
of the bosonic sector of gauged N=4, D=4
$\mathrm{SU(2)}\times\mathrm{SU(2)}$ supergravity. These configurations
generalize in a certain sense the Freedman-Gibbons solutions
\cite{freedgibb}, in that they
are not simply direct product spacetimes, but rather warped products.
Furthermore the dilaton is no more constant, it depends on the radial
coordinate $r$. The obtained spacetimes represent toroidal or
spherical black holes respectively naked singularities, depending on
the parameters. The spherical solutions break all
supersymmetries, whereas in the toroidal case we found BPS configurations
preserving one quarter of the supersymmetries, in particular,
we found an extremal BPS black hole. We further showed
that the background of the
toroidal spacetimes, which is obtained if the mass and the two electric
charges of the black hole are set to zero, breaks only half of the
supersymmetries, and corresponds to the domain wall solution of the
FS model.
In all cases, the Killing spinors depend only on the
radial coordinate $r$.\\
We have also used the Kaluza-Klein interpretation of gauged N=4, D=4
$\mathrm{SU(2)}\times\mathrm{SU(2)}$ supergravity to lift our
configurations to ten and eleven dimensions, so that they can be regarded
as solutions of the low energy effective action of type I- or
heterotic string-theory or of M-theory.
The lifted metrics result to have a rather complicated form,
with off-diagonal elements coming from the four-dimensional
gauge fields. However,
for $r \to \infty$ the ten-dimensional line element asymptotes in
string frame to a simple product spacetime
$M^2 \times S^1 \times S^1 \times S^3 \times S^3$ (endowed
with standard metric), which is equivalent to the near-horizon limit of
the intersection of two NS five-branes on a line.
It would be very tempting to try to use the
powerful tools of string- and M-theory in order to give a microscopic
interpretation of the Bekenstein-Hawking entropy of the four-dimensional
extremal toroidal BPS black hole.\\
As it is well known that N=2, D=7 gauged supergravity can be obtained
by compactifying N=1, D=10 supergravity on a three-sphere
(cf.~e.~g.~\cite{cowtown}), the solutions
found in our paper also give rise to new supersymmetric black holes in the
$\mathrm{SU(2)}$ gauged seven-dimensional theory.

\section*{Acknowledgement}

This work has been supported by a research grant within the
scope of the {\em Common Special Academic Program III} of the
Federal Republic of Germany and its Federal States, mediated 
by the DAAD.

\end{document}